\documentstyle[preprint,prb,aps,]{revtex}

\begin{document}
\draft
\title{Pseudogap formation of four-layer BaRuO$_3$ and its electrodynamic response
changes}
\author{Y. S. Lee, J. S. Lee, K. W. Kim, and T. W. Noh}
\address{School of Physics and Research Center for Oxide Electronics, Seoul National\\
University, Seoul 151-747, Korea}
\author{Jaejun Yu}
\address{School of Physics and Center for Strongly Correlated Materials Research,\\
Seoul National University, Seoul 151-747, Korea}
\author{Yunkyu Bang}
\address{Department of Physics, Chonnam National University, Kwangju 500-757, Korea}
\author{M. K. Lee and C. B. Eom}
\address{Department of Material Science and Engineering, University of\\
Wisconsin-Madison, Madison, Wisconsin 53706, U.S.A.}
\date{\today }
\maketitle

\begin{abstract}
We investiaged the optical properties of four-layer BaRuO$_{3}$, which shows
a fermi-liquid-like behavior at low temperature. Its optical conductivity
spectra clearly displayed the formation of a pseudogap and the development
of a coherent peak with decreasing temperature. Temperature-dependences of
the density $n$ and the scattering rate $1/\tau $ of the coherent component
were also derived. As the temperature decreases, both $n$ and $1/\tau $
decrease for four-layer BaRuO$_{3}$. These electrodynamic responses were
compared with those of nine-layer BaRuO$_{3}$, which also shows a pseudogap
formation but has an insulator-like state at low temperature. It was found
that the relative rates of change of both $n$ and $1/\tau $ determine either
metallic or insulator-like responses in the ruthenates. The optical
properties of the four-layer ruthenate were also compared with those of
other pseudogap systems, such as high $T_{c}$ cuprates and heavy electron
systems.
\end{abstract}


\newpage

\section{Introduction}

Recently, ruthenates have attracted a great deal of attention due to their
physical properties, exhibiting many unusual characteristics of strongly
correlated electron systems, such as unconventional superconductivity in
layered perovskite Sr$_{2}$RuO$_{4}$\cite{Maeno1994} and bad metallic
non-fermi-liquid behaviors in perovskite SrRuO$_{3}$.\cite{schlesinger98}
Although the ruthenate systems belong to 4$d$ transition metal oxides,
electron-electron interactions are expected to be important in determining
their physical properties. More recently, we reported a pseudogap formation
in the nine-layer rhombohedral (9{\it R}) BaRuO$_{3}$, which can be
considered as a new kind of pseudogap system.\cite{yslee1}

Pseudogap phenomena have been intensively investigated in some strongly
correlated electron systems. In high $T_{c}$ superconductors (HTSC), which
are 3$d$ transition metal oxides, pseudogap features with a reduction in the
density-of-state at the Fermi energy $N(E_{F})$ were observed especially in
the underdoped regime. These features have attracted lots of interests due
to their possible connection to the mechanism for the high $T_{c}$
superconductivity.\cite{pseudogap2,ARPES,HTSC1} In some $f$-electron
systems, such as UPt$_{3}$\cite{Heavyfermion} and UPd$_{2}$Al$_{3}$,\cite
{Heavyfermion3} pseudogaps were observed as well as heavy fermion (HF)
behaviors, which are due to the large enhancement of the effective mass $%
m^{\ast }$ of quasiparticles.\cite{Heavyfermion2} In both systems, the
scattering rate $1/\tau $ of quasiparticles was found to be suppressed with
pseudogap formation. Although many investigations have been performed, the
basic mechanisms of pseudogap formation have not yet been fully understood.

Our early observation of a pseudogap in 9{\it R} BaRuO$_{3}$ suggests that
the ruthenate could be another material with fascinating electronic
properties. \cite{yslee1} Compared to the cases of HTSC and HF systems, our
optical spectra show clearer electrodynamic response changes with pseudogap
formation. It was found that 9{\it R} BaRuO$_{3}$ can have a metallic state
in spite of a large reduction of carrier density $n$ after the pseudogap was
formed. This unusual behavior is found to be closely related to a very
significant reduction in $1/\tau $. It was also suggested that the pseudogap
formation could be related to structural characteristics of the layered
compound, which is shown in Fig. 1(a).\cite{donohue1964} In the 9{\it R}
structure, three adjacent RuO$_{6}$ octahedra have face-sharings, which
result in quasi-one-dimensional (1D) metal-bondings along the $c$-axis.
Then, a 1D-like instability fluctuation, such as a charge density wave (CDW)
instability, could contribute to the pseudogap formation.

To obtain further understanding, it would be useful to investigate
electrodynamic responses of ruthenates with similar crystal structures. Bulk
BaRuO$_{3}$ is known to have three crystal structures\cite
{donohue1964,6h,hong97}: (i) the 9{\it R}, (ii) the six-layered hexagonal (6%
{\it H}), and (iii) the four-layered hexagonal (4{\it H}) structures. The 9%
{\it R} BaRuO$_{3}$ phase has been considered the most stable bulk form. On
the other hand, 6{\it H} and 4{\it H} BaRuO$_{3}$ are known to be difficult
to synthesize reproducibly under ordinary conditions. Recently, some of us
(M. K. Lee and C. B. Eom) successfully synthesized epitaxial 4{\it H} BaRuO$%
_{3}$ films.\cite{Eom00,Lee00} As shown in Fig. 1(b), 4{\it H} BaRuO$_{3}$
has two adjacent RuO$_{6}$ octahedra with a face-sharing, so the quasi-1D
metal bonding can also occur along the $c$-axis. However, the 1D-like
contribution in the 4{\it H} compound will be weaker than that in 9{\it R}
BaRuO$_{3}$.

In spite of the structural similarities, some physical properties of these
layered compounds are quite different.\cite{Cava99} While the {\it dc}
resistivity $\rho (T)$ curve of 9{\it R} BaRuO$_{3}$ shows a crossover
behavior from a metallic to an insulator-like state at a low temperature ($T$%
), 4{\it H} BaRuO$_{3}$ remains in a metallic state down to very low $T$. It
is known that BaRuO$_{3}$ systems are non-magnetic and are expected to have
Pauli paramagnetism. As $T$ decreases, the magnetic susceptibility $\chi (T)$
of 4{\it H} BaRuO$_{3}$ increases slightly, however, that of 9{\it R} BaRuO$%
_{3}$ decreases.\cite{Cava99} These differences in the physical properties
can originate from subtle differences in the $T$-dependence of
electrodynamic quantities, such as $n$, $1/\tau $ and $m^{\ast }$. However,
there are no systematic studies on these intriguing quantities of 4{\it H}
BaRuO$_{3}$.

In this paper, we report the optical properties of 4{\it H} BaRuO$_{3}$. Its
optical conductivity spectra $\sigma _{1}(\omega )$ show pseudogap formation
and the development of a coherent component in the far-infrared region. With
the pseudogap formation, $n$ decreases and $1/\tau $ becomes suppressed. The
stronger $T$-dependence of $1/\tau $ will make this material metallic down
to very low $T$. From these observations, we suggest that the 1D-like
instability fluctuation could also be involved in 4{\it H} BaRuO$_{3}$, but
weaker than that in 9{\it R} BaRuO$_{3}$. The $T$-dependences of the
electrodynamic quantities will also be discussed in order to explain the
differences in $\rho (T)$ and $\chi (T)$ in the layered BaRuO$_{3}$
compounds.

\section{Experimental results}

\subsection{Sample preparation and characterization}

A high quality 4{\it H} BaRuO$_{3}$ epitaxial film was fabricated on a SrTiO$%
_{3}$(111) substrate using a 90$^{\text{o}}$ off-axis sputtering technique. 
\cite{Eom00} X-ray diffraction and transmission electron microscopy studies
revealed that the film is composed of single domains of {\it c}-axis 4{\it H}
structures with an in-plane epitaxial arrangement of BaRuO$_{3}$[2110]$\Vert 
$SrTiO$_{3}$[110]. Most measurements reported in this paper were performed
to probe properties parallel to the film, namely $ab$-plane responses of 4%
{\it H} BaRuO$_{3}$.

The $\rho (T)$ curve was measured using the standard four probe method. As
shown in Fig. 2, 4{\it H} BaRuO$_{3}$ has a metallic state down to 4 K.
Above 220 K, $\rho (T)$ seems to have a linear $T$-dependence, which has
been often observed in other bad metallic materials, such as SrRuO$_{3}$. In
4{\it H} BaRuO$_{3}$, the room temperature resistivity is comparable to the
Mott critical resistivity $\rho _{Mott}=3\hbar a/e^{2}\simeq 660$ $\Omega $%
cm, where {\it a} is a lattice constant of the $ab$-plane. So, the linear $T$%
-dependence of $\rho (T)$ near the room temperature region might be related
to the behavior of a bad metal. On the other hand, as shown in the inset of
Fig. 2, the $T$-dependence of $\rho (T)$ at low $T$ can scale with $T^{2}$,
indicating that 4{\it H} BaRuO$_{3}$ has a Fermi liquid-like behavior. This
strong Fermi liquid behavior can be evidently distinguished from the
insulator-like behavior of 9{\it R} BaRuO$_{3}$ below 110 K.

\subsection{Reflectivity measurements}

Near-normal incident reflectivity spectra, $R(\omega )$, of the {\it ab}%
-plane were measured in a wide photon energy region of 5 meV $\sim $ 30 eV
with $T$ variations. We used a conventional Fourier transform
spectrophotometer between 5 meV and 0.6 eV. Between 0.5 eV and 6.0 eV, we
used grating spectrometers. Above 6.0 eV, we used the synchrotron radiation
from the normal incidence monochromator beam line at Pohang Accelerator
Laboratory. To eliminate measurement errors due to the reflected light from
the film/substrate interface, we used a very thick film whose thickness was
about 1.1 $\mu $m. Then, the penetration depth, $\delta \sim c/\sqrt{2\pi
\omega \sigma }$, of a light at 100 cm$^{-1}$ was estimated to be about $0.6$
$\mu $m at 300 K. Since the film thickness is larger than $\delta $, the
contribution of the reflected light from the film/substrate interface could
be negligible. So, the strong phonon absorption peaks due to the SrTiO$_{3}$
substrate could not be observed in $R(\omega )$.

Figure 3 shows $T$-dependent $R(\omega )$ of 4{\it H} BaRuO$_{3}$ in the
far-infrared (IR) region. The inset of Fig. 3 shows $R(\omega )$ of 4{\it H}
BaRuO$_{3}$ in a wider frequency region. The level of $R(\omega )$ at 300 K
is high in the far-IR region in agreement with the fact that it is in a
metallic state. In addition, $R(\omega )$ below 200 cm$^{-1}$ are highly
consistent with predictions of the Hagen-Rubens relation, $R\simeq
1-(2\omega \rho /\pi )^{1/2}$.\cite{wooten} However, $R(\omega )$ show an
interesting crossover behavior: as $T$ decreases, the reflectivity becomes
strongly suppressed in the region between 200 cm$^{-1}$ and 500 cm$^{-1}$.
This crossover behavior is different from a typical metallic response,
indicating that 4{\it H} BaRuO$_{3}$ could have an unusual metallic state.

\section{Data analysis and results}

\subsection{Optical conductivity spectra}

The Kramers-Kronig (K-K) transformation was used to calculate complex
optical conductivity spectra, $\widetilde{\sigma }(\omega )\equiv \sigma
_{1}(\omega )-i\frac{\omega }{4\pi }\cdot \varepsilon _{1}(\omega )$, from
the experimental $R(\omega )$. Here, $\varepsilon _{1}(\omega )$ represent
real dielectric function spectra. For the K-K transformation, $R(\omega )$
in the low frequency region were extrapolated with the Hagen-Rubens
relation. \cite{wooten} The calculated $\sigma _{1}(\omega )$ were found to
be consistent with the experimental $\sigma _{1}(\omega )$ independently
obtained by spectroscopic ellipsometry techniques in the visible region.
These consistencies demonstrate the validity of our K-K analysis.\cite
{jsahn99} The details of this analysis are published elsewhere.\cite{hschoi}

Figure 4 shows $T$-dependent $\sigma _{1}(\omega )$ obtained through the K-K
analysis in a wide frequency region. The strong peak at 550 cm$^{-1}$ and
several weak peaks below 500 cm$^{-1}$ are due to transverse optic phonon
modes. At 300 K, $\sigma _{1}(\omega )$ seems to show a Drude-like behavior
with a very large value of 1/$\tau $. As $T$ decreases, however, the
spectral weight below 500 cm$^{-1}$ becomes suppressed and the spectral
weight near 800 cm$^{-1}$ increases, forming a gap-like feature. This
feature is different from the formation of a normal energy gap in an
insulator (or a semiconductor), where the spectral weight in the gap-like
region should vanish completely. The remainder of the spectral weights in
the gap-like region indicates that a pseudogap is formed in this material.

Figure 5(a) shows the pseudogap formation of 4{\it H} BaRuO$_{3}$ in more
detail. One of the most important experimental facts is that the sum rule is
satisfied: when $\sigma _{1}(\omega )$ is integrated up to 1.0 eV, the
optical strength remains nearly constant. This means that most of the
missing spectral weights in the gap-like region are transferred to higher
frequencies, especially near 650 cm$^{-1}\simeq 0.08$ eV. [From now on, let
us call the peak frequency position of $\sigma _{1}(\omega )$ the pseudogap
position, $\omega _{c}$.] Another interesting fact is that most of the
spectral weight gain occurs near $\omega _{c}$. Above 1,400 cm$^{-1}$, the $%
T $-dependence of the spectral weight changes is quite small. The
significant spectral weight gain just above $\omega _{c}$ is also different
from that of the normal energy gap formation, where the missed spectral
weight is redistributed in much wider frequency regions.

Note that the pseudogap is formed in the metallic region, judged from $\rho
(T)$. This fact is also related to a clear development of a strong
Drude-like component in the far-IR region, as shown in Fig 5(b). As $T$
decreases, $\sigma _{1}(\omega )$ show that a coherent Drude-like component
becomes more evident: the value of $\sigma _{1}(\omega )$ in the low
frequency region becomes larger and the width of the coherent component
becomes smaller. The concurrent development of the pseudogap and coherent
peak was also observed in 9{\it R} BaRuO$_{3}$,\cite{yslee1} however, such
phenomena are more clearly seen in the 4{\it H} compound.

\subsection{Inapplicability of the single component analyses}

The optical responses of free carriers are usually analyzed in terms of the
simple Drude model:

\begin{equation}
\widetilde{\sigma }(\omega )\equiv \sigma _{1}(\omega )-i\frac{\omega }{4\pi 
}\cdot \varepsilon _{1}(\omega )=\frac{\omega _{p}^{2}}{4\pi }\cdot \frac{%
\tau }{(1-i\tau \omega )}\text{,}
\end{equation}
where the plasma frequency, $\omega _{p}=\sqrt{4\pi ne^{2}/m^{\ast }}$,
represents the spectral weight of free carriers. However, we cannot apply
this formula to explain the measured $\widetilde{\sigma }(\omega )$ up to
1.0 eV. The room temperature data, shown in Fig. 4, might be approximately
explained in terms of this formula. Then, the value of 1/$\tau $ could be
estimated to be about 4,500 cm$^{-1}$. This value is so large that it seems
difficult to explain the metallic state of this material. On the other hand,
as shown in Fig. 5(b), the apparent 1/$\tau $ at 10 K for the coherent
component falls below 100 cm$^{-1}$, much lower than the estimated value at
room temperature. This suggests that all of the $T$-dependent $\widetilde{%
\sigma }(\omega )$ up to 1.0 eV cannot be explained in terms of the simple
Drude model.

In many strongly correlated systems, optical responses of quasiparticles
have been explained in terms of the frequency-dependent scattering rate $%
1/\tau (\omega )$ and effective mass enhancement $m^{\ast }(\omega )$.\cite
{edm1} Under such assumptions, the optical responses can be described in
terms of the extended Drude model:

\begin{equation}
\widetilde{\sigma }(\omega )=\frac{\omega _{p}^{2}}{4\pi }\cdot \frac{\tau
(\omega )}{[1-i\omega m^{\ast }(\omega )\tau (\omega )]}\text{.}
\end{equation}
By using Eq. (2), both $m^{\ast }(\omega )$ and $1/\tau (\omega )$ can be
derived directly from the experimental $\widetilde{\sigma }(\omega )$. For
most pseudogap materials, such as HTSC and HF systems, their optical
responses have been interpreted in terms of this model.\cite
{Heavyfermion2,Ginsberg} However, we found that it was difficult to extend
such single component analysis to the 4{\it H} BaRuO$_{3}$ case. Figure 6
(a) and (b) show $1/\tau (\omega )$ and $m^{\ast }(\omega )$ obtained from
an analysis based on the extended Drude model, respectively. Near $\omega
_{c}$, both $1/\tau (\omega )$ and $m^{\ast }(\omega )$ do not make sense.
Especially, $m^{\ast }(\omega )$ becomes negative in the frequency region
near $\omega _{c}$, which is inadequate from a physics point of view.
Therefore, the single component analysis based on the extended Drude model
cannot also be applied to this layered ruthenate system.

\subsection{$T$-dependence of electrodynamic quantities using the two
component analysis}

For quantitative analysis of the $T$-dependent coherent component, we fitted
the complex optical conductivity $\widetilde{\sigma }(\omega )$ up to 1.0 eV
using a two-component model. At each temperature, the coherent component in
the far-IR region and gap-like feature near $\omega _{c}$ were fitted with
the simple Drude model in Eq. (1) and the Lorentz oscillator model,
respectively. The fitting results for $1/\tau $ and $\omega _{p}^{2}$ of the
coherent components are shown in Fig. 7.

It is noted that $\omega _{p}^{2}$, which is proportional to the spectral
weight of the coherent component, is reduced as the pseudogap develops. In
general, the reduction of $\omega _{p}^{2}$ can occur either by a reduction
of $n$ or by an enhancement of $m^{\ast }$. For approximate estimation of $n$
and $m^{\ast }$, we used the three-dimensional free electron gas model. In
this model, $\chi \propto m^{\ast }n^{1/3}$ and $\omega _{p}^{2}\propto
n/m^{\ast }$. With the previous reported values of $\chi (T)$, \cite{Cava99}
the free electron gas model provides $n$ $\simeq 8.3\times 10^{21}$ cm$^{-3}$
and $m^{\ast }/m_{e}\simeq 36$ at 10 K for 4{\it H} BaRuO$_{3}$, so its mean
free path $\ell $ can be estimated to be about 19 \r{A}. At 300 K, the model
provides $n$ $\simeq 1.2\times 10^{22}$ cm$^{-3}$ and $m^{\ast }/m_{e}\simeq
27$, where $m_{e}$ is a bare electron mass, so $\ell $ will be about 4 \r{A}%
. \cite{comment2} This value is comparable to the lattice constant of the $%
ab $-plane, i.e., about 5.7 \r{A}, which could be related to its bad
metallic behavior, mentioned in Sect. II A.

In spite of the reduction of $n$, 4{\it H} BaRuO$_{3}$ has a metallic state
down to low $T$. This unusual behavior can be explained by $T$-dependent
changes in $\omega _{p}^{2}$ and 1/$\tau $. As shown in Fig. 7, $1/\tau $
might decrease significantly. According to the Drude model, {\it dc}
electric resistivity $\rho _{dc}$ is inversely proportional to $\omega
_{p}^{2}$ and $\tau $. So, the relative changes in $\omega _{p}^{2}$ and 1/$%
\tau $ are important for determining a metallic or an insulator-like
behavior in $\rho _{dc}$. If the decrease in 1/$\tau $ is larger than the
reduction of $n$, a metallic state can be induced. The strong $T$-dependent
rapid decrease in 1/$\tau $ can explain the metallic behavior of 4{\it H}
BaRuO$_{3}$, which might be closely related to the origin of the pseudogap.

\section{Discussion}

\subsection{Fermi liquid behaviors of 4{\it H} BaRuO$_3$}

Contrary to the 9{\it R} compound, 4{\it H} BaRuO$_{3}$ shows a Fermi liquid
behavior in a low $T$ region. As shown in the inset of Fig. 2, its $T$%
-dependence of $\rho (T)$ at low $T$ can scale as $T^{2}$, indicating that 4%
{\it H} BaRuO$_{3}$ has Fermi liquid behavior. The fact that the $T^{2}$%
-scaling is valid in a relatively large $T$ region suggests that the
electron-electron interaction is quite strong in this layered ruthenate.\cite
{comment4} From the relation $\rho (T)=\rho _{o}+AT^{2}$, $A$ is found to be
about 2.9$\times $10$^{-2}$ $\mu \Omega $cm/K$^{2}$. This value is
comparable with that found by a previous report on the 4{\it H} BaRuO$_{3}$
single crystal.\cite{Cava99}

We can estimate a linear coefficient in the specific heat $\gamma $ to be
about $55$ mJ/mol K$^{2}$, using the universal relation, $A$/$\gamma
^{2}\simeq 1.0\times $10$^{-5}$ $\mu \Omega $cm (mol $\cdot $ K/mJ)$^{2}$. 
\cite{KWrelation} A similar value of $\gamma $ can be obtained from a
reported value of $\chi (T)\simeq 7\times 10^{-4}$ emu/cm$^{3}$ within the
standard free-electron model.\cite{Cava99} It is interesting to note that
these values are comparable with those of Sr$_{2}$RuO$_{4}$ with a
superconducting ground state [$\chi \simeq 9.7\times 10^{-4}$ emu/mol and $%
\gamma \simeq 39$ mJ/mol K$^{2}$ for K].\cite{Maeno1994} Compared with the
corresponding values for a typical Pauli paramagnetic metal (e.g. $\gamma
\simeq 0.7$ mJ/mol K$^{2}$ for Au, $\chi \simeq 2.6\times 10^{-5}$ emu/mol
for K),\cite{text1} the values of $\gamma $ and $\chi $ might be greatly
enhanced. For Sr$_{2}$RuO$_{4}$, the value of $\chi (T)$ is significantly
about 15 times that expected from the standard free electron model with $%
n\simeq 2.0\times 10^{22}$ cm$^{-3}$ and $m^{\ast }\simeq 2m_{e}$. This
enhancement has been explained by ferromagnetic fluctuations.\cite{Sr214_2}
Judging from the case of Sr$_{2}$RuO$_{4}$, it is possible that the
ferromagnetic fluctuation might also be related to the large value of $\chi
(T)$ in 4{\it H} BaRuO$_{3}$.\cite{comment1} To clarify this possibility,
further investigations are necessary.

\subsection{Comparisons between 4{\it H} and 9{\it R} BaRuO$_{3}$}

The optical spectra of 4{\it H} and 9{\it R} BaRuO$_{3}$ show the pseudogap
formation and the development of the coherent components. As $T$ decreases,
both of the ruthenates show a reduction in $\omega _{p}^{2}$ and suppression
in 1/$\tau $. [Refer to Fig. 7 of this paper and Fig. 3 of Ref. 3] Although
the general trends in the $T$-dependences of the electrodynamic quantities
are similar, some of their transport and magnetic properties are different. 
\cite{Cava99} These differences should be explained in terms of subtle
differences of the $T$-dependences in the electrodynamic quantities.

While 4{\it H} BaRuO$_{3}$ remains in a metallic state down to a very low $T$%
, the $\rho (T)$ curve of 9{\it R} BaRuO$_{3}$ shows a crossover behavior
from a metallic (i.e. $d\rho /dT>0$) to an insulator-like state (i.e. $d\rho
/dT<0$) around 110 K.\cite{yslee1} Since $\rho _{dc}$ is inversely
proportional to $\omega _{p}^{2}$ and $\tau $, the relative changes in $%
\omega _{p}^{2}$ and 1/$\tau $ will determine a metallic or an
insulator-like state in $\rho (T)$. For 4{\it H} BaRuO$_{3}$, the decrease
in 1/$\tau $ is larger than that for $\omega _{p}^{2}$ at all temperatures,
as shown in Fig. 7. So, $d\rho /dT>0$ for all temperatures. On the other
hand, for 9{\it R} BaRuO$_{3}$, the reduction in 1/$\tau $ dominates and
leads to metallic behavior above 110 K. However, at low $T$, the reduction
in $\omega _{p}^{2}$ is quite considerable and the inelastic relaxation time
can become very large, so the disorder-induced scattering starts to play an
important role. Then, the decrease in 1/$\tau $ becomes smaller than that
for $\omega _{p}^{2}$, resulting in the insulator-like state of $d\rho /dT<0$%
.

As $T$ decreases, the magnetic susceptibility $\chi (T)$ of 4{\it H} BaRuO$%
_{3}$ increases slightly, however, that of 9{\it R} BaRuO$_{3}$ decreases. 
\cite{Cava99} It is known that both of the BaRuO$_{3}$ systems are
non-magnetic and are expected to have Pauli paramagnetism, whose
contribution is proportional to $m^{\ast }n^{1/3}$. Note that $\omega
_{p}^{2}\propto n/m$, and that the changes in $\omega _{p}^{2}$ of the 4{\it %
H} and 9{\it R} BaRuO$_{3}$ are about a factor of 2 and 30, respectively.
The larger change in $\omega _{p}^{2}$ of 9{\it R} BaRuO$_{3}$, which is due
to a significant reduction in $n$, results in a decrease in $\chi (T)$. On
the other hand, for 4{\it H} BaRuO$_{3}$, the subtle competition between
weakly $T$-dependent $n(T)$ and $m^{\ast }(T)$ can enhance $\chi (T)$
slightly.

\subsection{Comparisons with other pseudogap systems}

In the case of underdoped HTSC, $\sigma _{1}(\omega )$ in the $ab$-plane
shows only weak spectral weight changes, so it is not easy to address the
pseudogap formation based solely on $\sigma _{1}(\omega )$. However, when
the extended Drude model is used, $1/\tau (\omega )$ shows a strong
suppression in the low frequency region, which has been identified as a
signature of the pseudogap.\cite{Basov} On the contrary, $\sigma _{1}(\omega
)$ along the $c$-axis exhibits a suppression of the spectral weight in the
far-IR region, which can be easily interpreted as a feature of the
pseudogap. \cite{homes} However, it has been observed in the insulating
state. On the other hand, BaRuO$_{3}$ systems show the clear pseudogap
formation in $\sigma _{1}(\omega )$ of the metallic state. The features due
to the pseudogap are so distinct that the extended Drude model analysis
cannot be applied in the frequency region near $\omega _{c}$, which is
addressed in Sect. II B.

Pseudogap formations in $\sigma _{1}(\omega )$ have been also observed in
some HF systems, such as UPt$_{3}$\cite{Heavyfermion} and UPd$_{2}$Al$_{3}$. 
\cite{Heavyfermion3} However, the very small pseudogap size ($\sim $ 0.01
eV) and the extremely small spectral weight of the coherent mode hindered
detailed analysis in connection with development of the pseudogap. For HF
systems, the electrodynamic quantities were analyzed by the extended Drude
model, assuming no change in $n$ with the pseudogap formation. With
decreasing $T$, $1/\tau (\omega )$ is also suppressed and $m^{\ast }(\omega )
$ becomes enhanced significantly. The enhancement of $m^{\ast }(\omega )$ in
the far-IR region has been explained by a strong renormalization of
quasiparticles due to interactions with localized $f$-electrons,\cite
{Heavyfermion2} which could be the origin of HF behavior. On the other hand,
our BaRuO$_{3}$ is a non-magnetic material with no $f$-electrons. In
contrast to the case of HF systems, a reduction in $n$ is clearly observed
with the pseudogap formation, and the single component analysis based on the
extended Drude model is not adequate to describe the spectral weight
changes. As shown in Sec. III C, the two component model can describe its
spectral weight changes better. The comparison in the electrodynamic
quantities between BaRuO$_{3}$ and other pseudogap materials is summarized
in TABLE I.

\subsection{The origin of the pseudogaps in BaRuO$_3$ : CDW fluctuations}

In our earlier paper on 9{\it R} BaRuO$_{3}$, we argued that the origin of
the pseudogap in the ruthenate should be the CDW fluctuations. Due to the
quasi-1D metal-metal bondings along the $c$-axis, shown in Fig. 1, such 1D
instability fluctuations can make a contribution. There are a few other
experimental facts which support this suggestion. First, BaIrO$_{3}$,
isostructural with 9{\it R} BaRuO$_{3}$, shows a static CDW instability.\cite
{BaIrO3} Second, the pseudogap positions of BaRuO$_{3}$ systems are at about
0.1 eV, which is comparable with the CDW gap value of a static CDW system. 
\cite{CDW3} Third, the spectral weight change in Fig. 5(a) is quite similar
to those predicted for the density wave materials.\cite{SDW2,Tinkham} We
believe that the observation of the pseudogap in 4{\it H} BaRuO$_{3}$
provides further support for the CDW fluctuations as the origin of the
pseudogap in the layered ruthenates.\cite{comment3}

With such CDW fluctuations, there could be a partial gap-opening in the $k$%
-space on the Fermi surface, which could be closely related to the case of
pseudogap formation in HTSC. Simultaneously, the suppression of the
scattering channels on the Fermi surface could induce the reduction of $%
1/\tau $. Similar behaviors are observed in some density wave materials. 2%
{\it H}-TaSe$_{2}$, which is a 2D CDW material, shows a weak gap structure,
but retains its metallic characteristics across the CDW ordering transition
temperature $T_{CDW}$. The absence of the increase of $\rho (T)$ at $T_{CDW}$
was attributed to the partial-gap opening over a restricted region of the
Fermi surface,\cite{2D_CDW} and furthermore, the reduction in $1/\tau $ is
suggested to occur due to the collapse of a scattering channel with the
formation of the CDW state.\cite{2D_CDW2}

Although the metallic state of the BaRuO$_{3}$ systems is similar to the
case of other static density wave systems, the ruthenates have quite
distinct features. First, in both 4{\it H} and 9{\it R} BaRuO$_{3}$, no
static density wave formation has been observed even down to very low $T$. 
\cite{comment5} Second, the ruthenates do not show any strong anisotropy in
their physical properties.\cite{Cava99} Third, the spectral weight transfer
with the pseudogap formation is quite similar to that of the 1D-like CDW
ordering. This strong 1D-like pseudogap feature is also distinguished from
the 2D CDW materials such as 2{\it H}-TaSe$_{2}$.\cite{2D_CDW} Fourth, the
pseudogap positions show no significant $T$-dependent behavior, so it does
not follow the BCS-like behavior for the density wave gap. Finally, a strong 
$T$-dependent suppression of $1/\tau $ occurs with a reduction in $n$ with
pseudogap formation. These distinct features, which could be related to the
lack of the long-range ordering at any finite temperature, might be related
to the quantum critical point fluctuation near a metal-insulator transition. 
\cite{Sachdev} Further investigations to probe this possibility are highly
desirable.

\subsection{A systematic trend in the layered compounds, BaMO$_{3}$ (M=Ru or
Ir)}

In the layered compounds with the face-sharing of the Ru(or Ir)O$_{6}$
octahedra, it was suggested that the strength of the quasi 1D-like
metal-bonding along the {\it c}-axis could be a parameter which controls
their physical properties.\cite{yslee1} The metal-bonding contribution in 9%
{\it R} BaRuO$_{3}$ becomes larger than that in 4{\it H} BaRuO$_{3}$, since
the former has greater numbers of face-sharings per single RuO$_{6}$
octahedron than the latter. The quasi 1D-like contribution will become even
larger for BaIrO$_{3}$, since the 5$d$ transition metal Ir ions could have a
stronger metallic bonding character than the 4$d$ transition metal Ru ions. 
\cite{BaIrO3}

It is quite interesting to see that there is a systematic trend in the BaMO$%
_{3}$ (M=Ru or Ir) compounds. From the optical spectra, we approximately
estimated the pseudogap positions to be about $0.08$ eV and $0.1$ eV in 4%
{\it H} and 9{\it R} BaRuO$_{3}$, respectively. It was also reported that
the optical gap position value of BaIrO$_{3}$ is about 0.15 eV.\cite{BaIrO3}
As the quasi-1D-like metal-bonding along $c$-axis becomes stronger, the
pseudogap (or gap) position in the $ab$-plane seems to become higher.
Moreover, the bonding strength seems to also be related to the transport
properties. 4{\it H} BaRuO$_{3}$ has a metallic behavior even with a small
reduction in $n$ at a low $T$. 9{\it R} BaRuO$_{3}$ shows an interesting
crossover from metallic to insulating-like states due to the significant
reduction in $n$ with decreasing $T$. BaIrO$_{3}$ seems to be an insulator
(namely, $n=0$). A summary of these systematic changes is given in TABLE II.
To confirm these interesting properties, it is useful to investigate further
the physical properties of other 4$d$ or 5$d$ compounds with 4{\it H}, 6{\it %
H}, and 9{\it R} structures.

\section{Summary}

4{\it H} BaRuO$_{3}$ exhibits clear pseudogap formation in its optical
spectra, similar to 9{\it R} BaRuO$_{3}$. From these observation, it can be
concluded that 4{\it H} and 9{\it R} BaRuO$_{3}$ compounds can be a new
class of pseudogap systems in 4{\it d} electron systems. It looks like that
the strong suppression of the scattering rate, in spite of the carrier
density, is responsible for their metallic states. Additionally, the subtle
difference in electrodynamic parameters makes two BaRuO$_{3}$ systems have
different temperature dependences of resistivity and magnetic
susceptibility. It is expected that these materials will provide more
insights on pseudogap phenomena in strongly correlated materials.

\acknowledgments
We acknowledge H. Y. Choi and K. H. Kim for helpful discussion. We also
thank the Inter-University Center for National Science Facilities at SNU and
Pohang Accelerator Laboratory for allowing us to use some of their
facilities. This work was supported by the Ministry of Science and
Technology through the Creative Research Initiative program.

\newpage

\begin{figure}[tbp]
\caption{Schematic diagrams of two crystallographic forms of BaRuO$_{3}$;
(a) 9{\it R} phase, (b) 4{\it H} phase. }
\label{Structural diagram}
\end{figure}

\begin{figure}[tbp]
\caption{$T$-dependent $\protect\rho (T)$ curve. The inset shows the linear $%
T^{2}$-dependence of $\protect\rho (T)$ curve at low $T$. The dotted line
represents a linear $T$-dependence at high $T$.}
\label{resitivity}
\end{figure}

\begin{figure}[tbp]
\caption{$T$-dependent $R(\protect\omega )$ below 1000 cm$^{-1}$. In the
inset, $R(\protect\omega )$ at 300 K in a wide frequency range is shown.}
\label{Reflectivity}
\end{figure}

\begin{figure}[tbp]
\caption{$T$-dependent $\protect\sigma _{1}(\protect\omega )$ of 4{\it H}
BaRuO$_{3}$ with a logarithmical scale of the bottom-axis. The dotted lines
are $\protect\sigma _{1}(\protect\omega )$ obtained from the Hagen-Rubens
extrapolations. The arrow represents $\protect\omega _{c}$, the pseudogap
position.}
\label{conductivity 2}
\end{figure}

\begin{figure}[tbp]
\caption{(a) $T$-dependenct $\protect\sigma _{1}(\protect\omega )$ below
3000 cm$^{-1}$. $\protect\omega _{c}$ represents the pseudogap value. (b) $%
\protect\sigma _{1}(\protect\omega )$ below 500 cm$^{-1}$ shown in detail.
The solid triangle, solid circle, and solid square represent the {\it dc} $%
\protect\sigma (T)$ at 80, 150, and 300 K, respectively. The dotted lines
are $\protect\sigma _{1}(\protect\omega )$ obtained from Hagen-Rubens
extrapolation.}
\label{conductivity 1}
\end{figure}

\begin{figure}[tbp]
\caption{$T$-dependent (a) $1/\protect\tau (\protect\omega )$ and (b) $m^{*}(%
\protect\omega )$ derived from the extended Drude model analysis with $%
\protect\omega _p \simeq$ 2.3 eV.}
\label{extended Drude model}
\end{figure}

\begin{figure}[tbp]
\caption{$T$-dependent $\protect\omega _p^2$ and $1/\protect\tau $ of the
coherent component in 4{\it H} BaRuO$_3$.}
\label{fitting result}
\end{figure}

\newpage

\begin{table}[tbp]
\caption{Summary of the changes of electrodynamic quantities with pseudogap
formation in the various pseudogap systems. [$^{1)}$ Ref. 29 and $^{2)}$
Ref. 9 \& 30.]}
\begin{tabular}[t]{c|ccc}
& HTSC$^{1)}$ & HF systems$^{2)}$ & BaRuO$_{3}$ \\ 
\tableline Anisotropy & strong & weak & weak \\ 
$n$ & constant & constant & decrease \\ 
$1/\tau $ & suppressed & strongly suppressed & strongly suppressed \\ 
$m^{\ast }$ & nearly constant & largely enhanced & nearly constant \\ 
\tableline Pseudogap origin &  & hybridization & 1D CDW fluctuation \\ 
Pseudogap size (eV) & $\sim $ 0.1 & $\sim $ 0.01 & $\sim $ 0.1 \\ 
Pseudgap feature in $\sigma _{1}(\omega )$ & weak & strong & strong
\end{tabular}
\end{table}

\begin{table}[tbp]
\caption{Summary of the physical properties in the layered materials (BaMO$%
_{3}$, M=Ru and Ir). [$^{1)}$ Ref. 3 and $^{2)}$ Ref. 31.]}
\begin{tabular}[t]{c|ccc}
& 4{\it H} BaRuO$_{3}$ & 9{\it R} BaRuO$_{3}^{1)}$ & BaIrO$_{3}^{2)}$ \\ 
\tableline\tableline Metal-bonding structure & two RuO$_{6}$ & three RuO$%
_{6} $ & three IrO$_{6}$ \\ 
\tableline(Pseudo) Gap positon & 0.08 eV & 0.1 eV & 0.15 eV \\ 
\tableline Electric state & Fermi-liquid-like & Crossover behavior & 
Insulating \\ 
at low $T$ &  & from metallic to insulator-like states & 
\end{tabular}
\end{table}


\begin{references}
\bibitem{comment2}  In this analysis, the vaules of $m^{*}$ are estimated to
be too large. This is because $\chi (T)$ in 4{\it H} BaRuO$_3$ might be
largely enhanced due to the ferromagnetic fluctuation.
\bibitem{comment4}  In this temperature region, the very weak $T$%
-dependences of $n(T)$ and $m^{*}(T)$ are observed. So, the $T^2$-dependence
is expected to come from electron-electron scatterings.
\end{references}
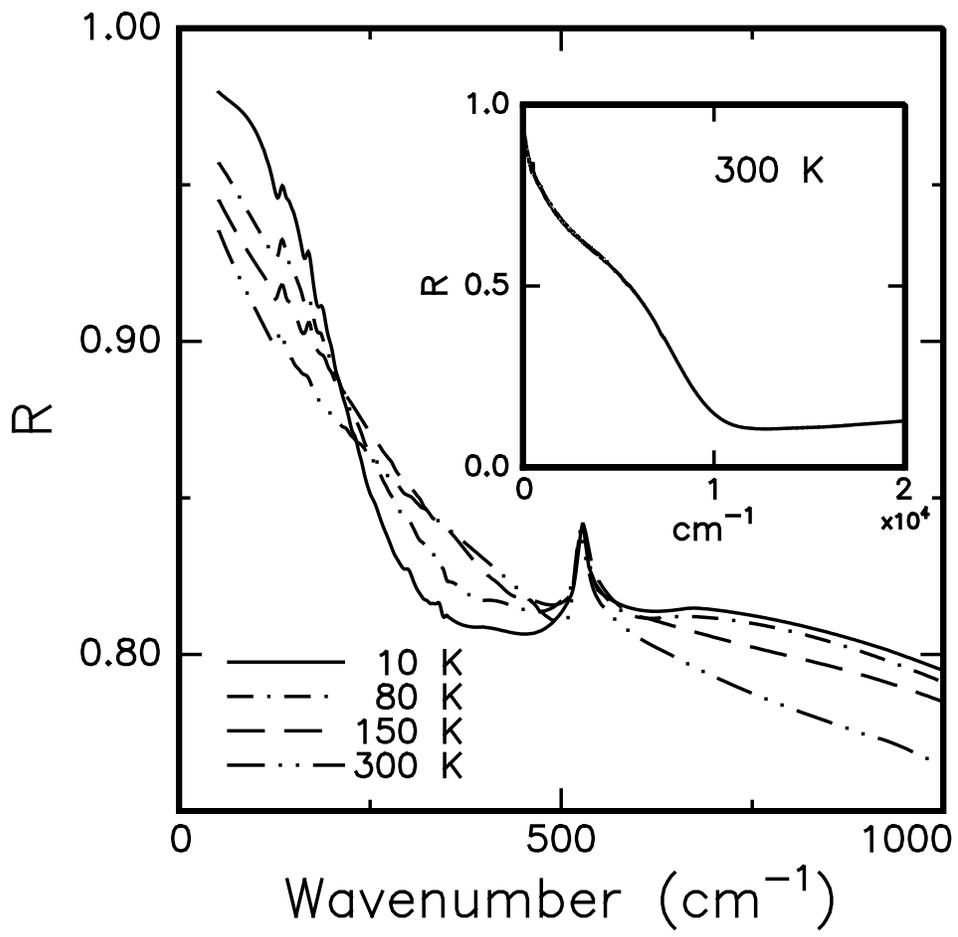
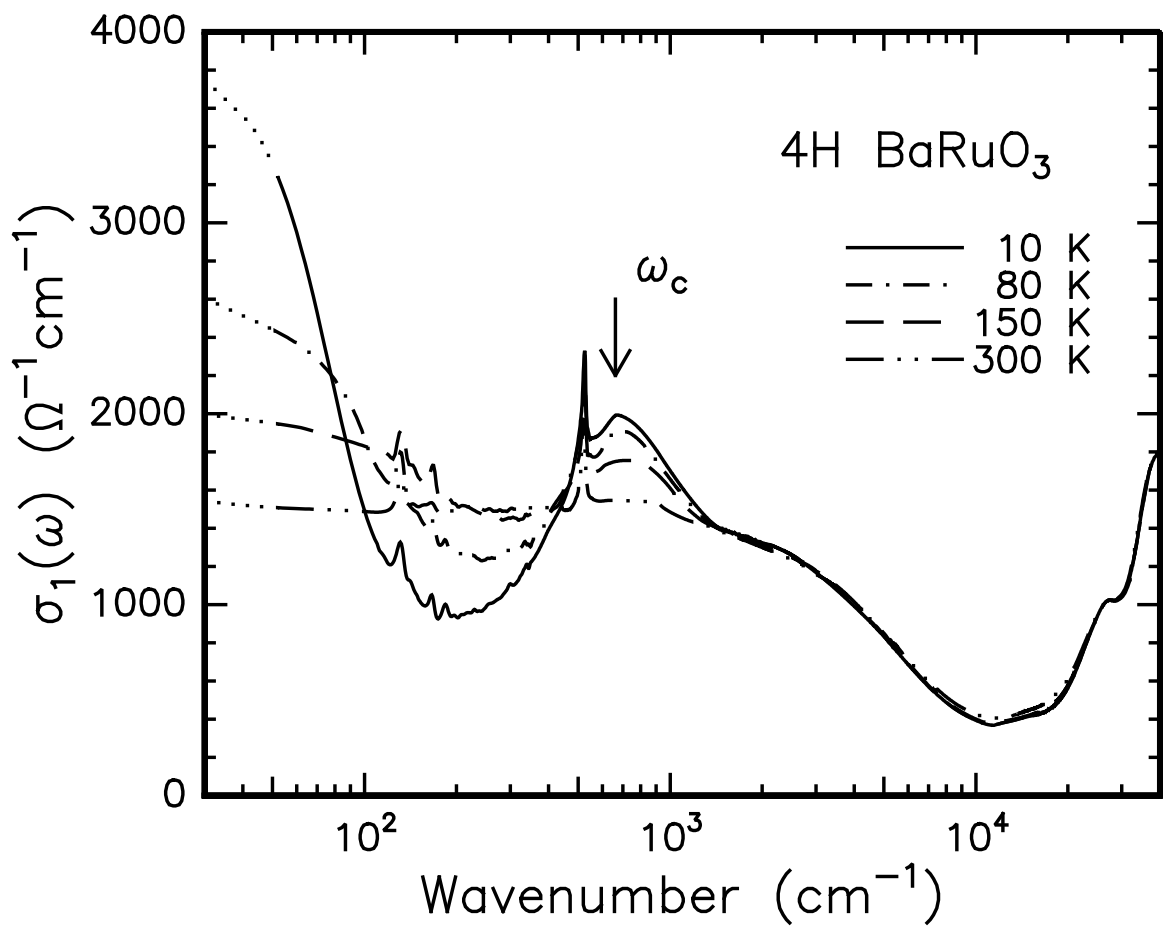
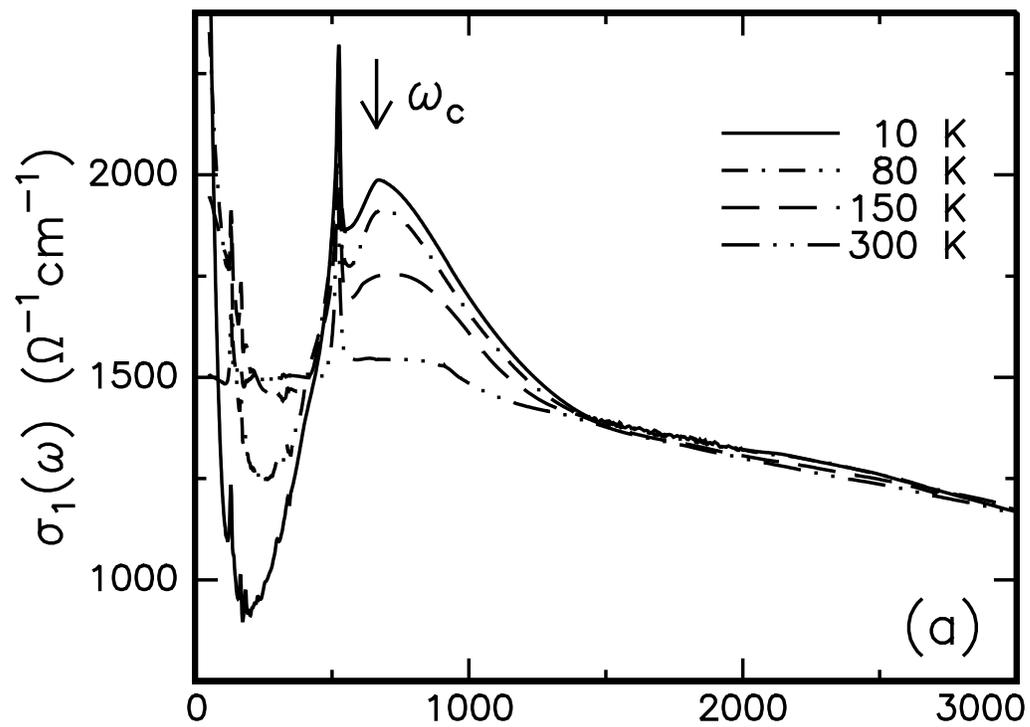
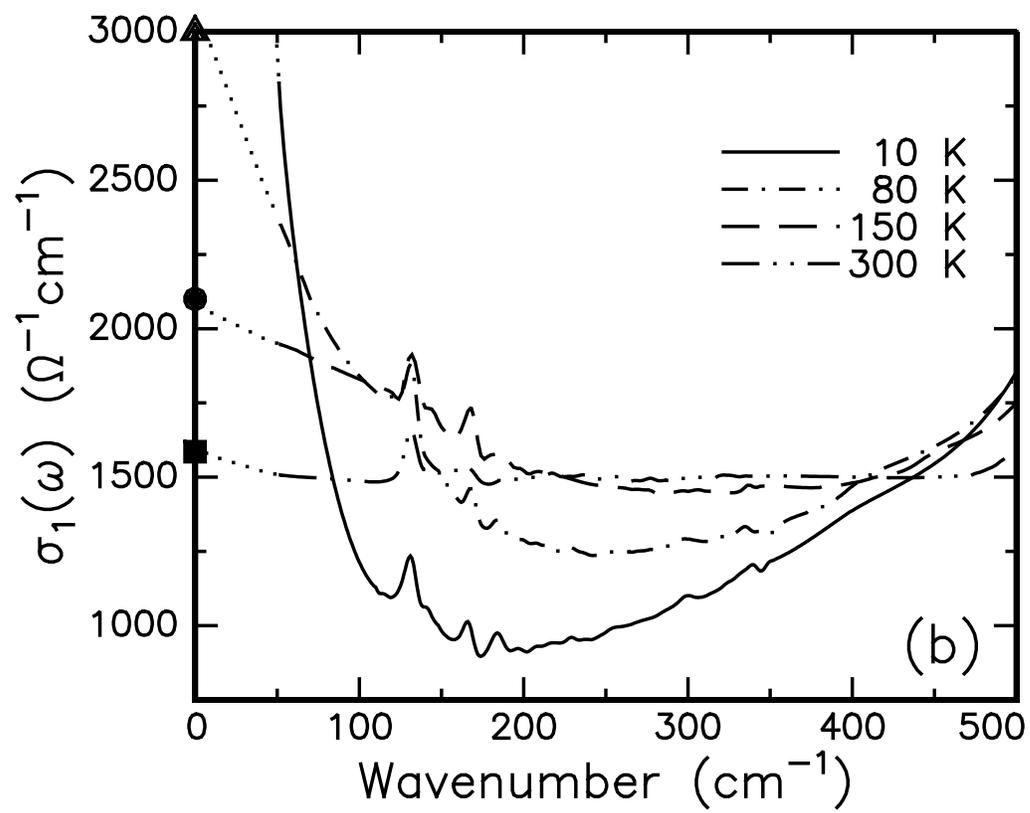
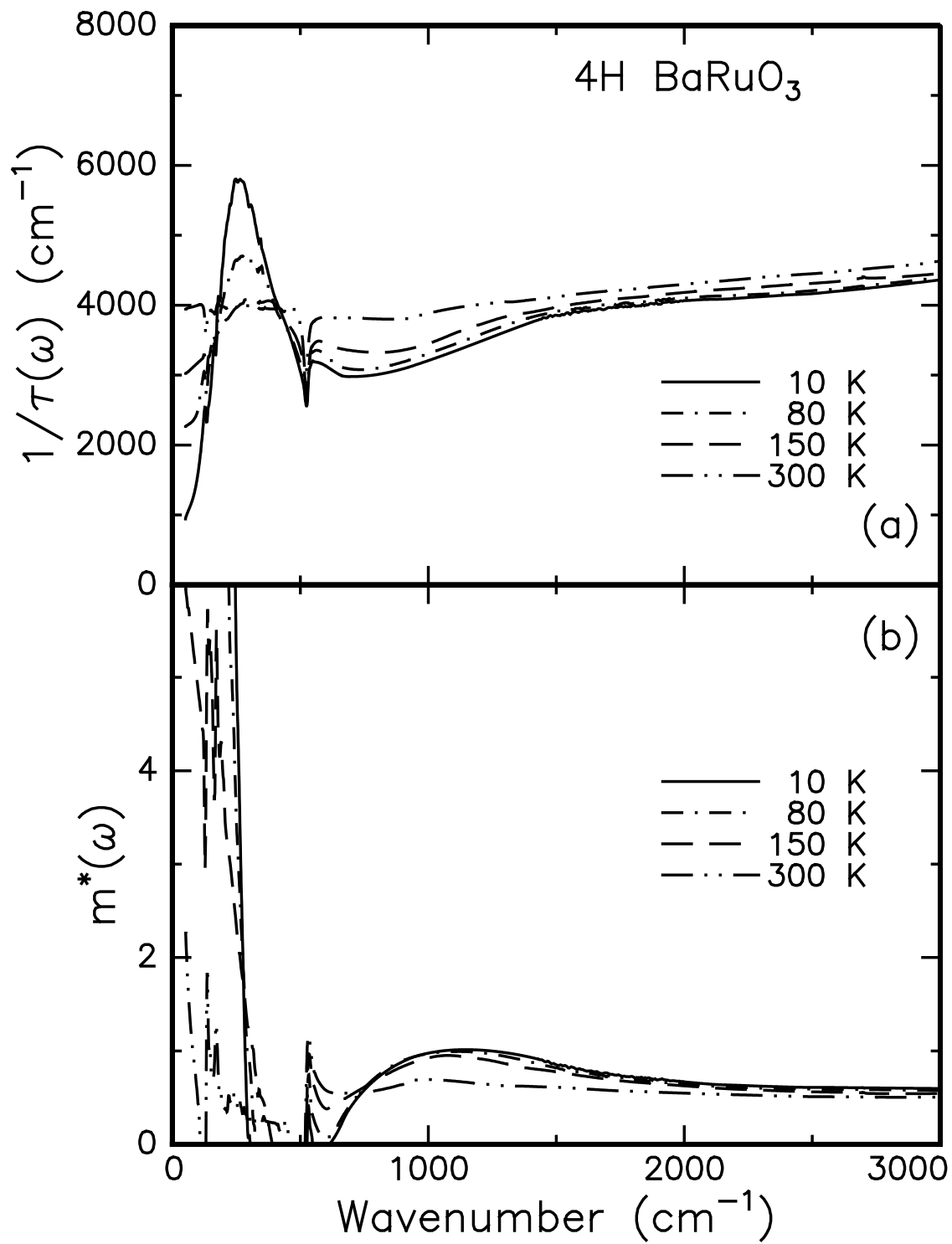
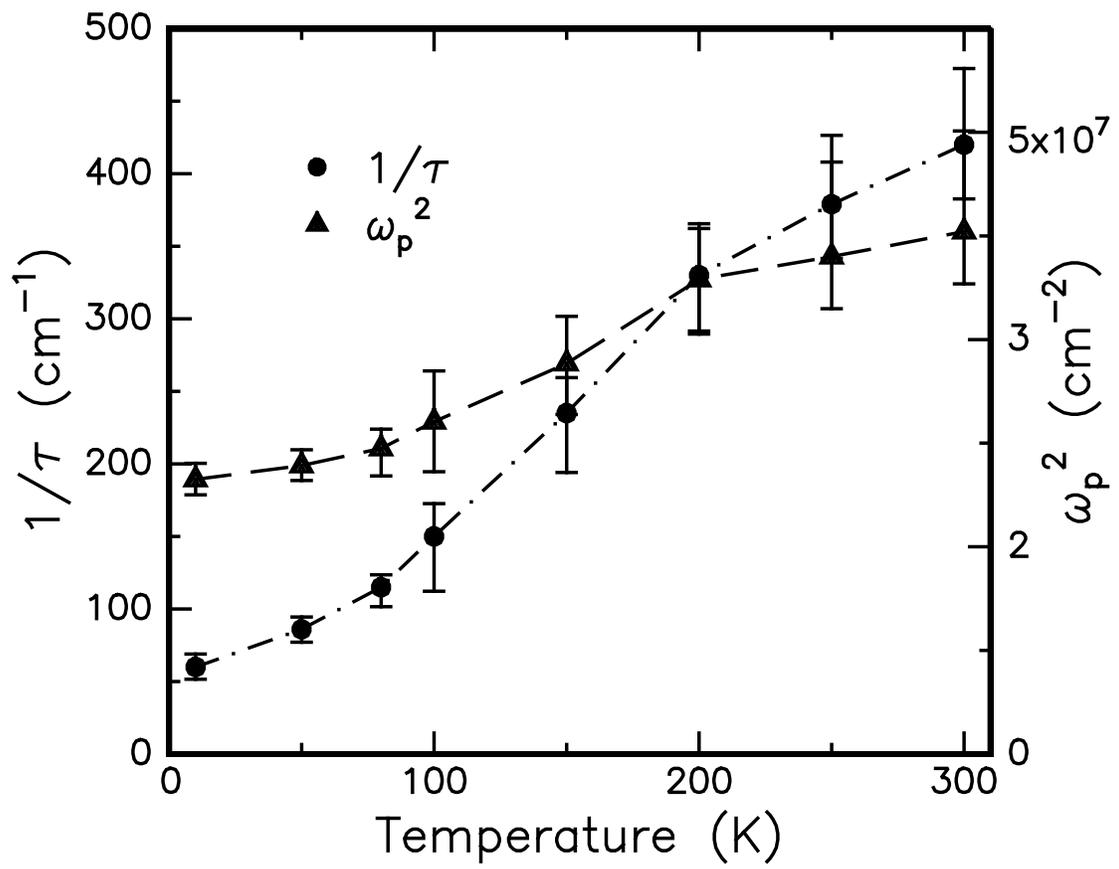
\end{document}